# Shannon Information Entropy in Position Space for Two-Electron Atomic Systems

Chien-Hao Lin[1#] and Yew Kam Ho[1*]

[1]Institute of Atomic and Molecular Sciences, Academia Sinica, Taiwan

## ABSTRACT

Entropic measures provide analytic tools to help us understand correlation in quantum systems. In our previous work, we calculated linear entropy and von Neumann entropy as entanglement measures for the ground state and lower lying excited states in helium-like systems. In this work, we adopt another entropic measure, Shannon entropy, to probe the nature of correlation effects. Besides the results of the Shannon entropy in coordinate space for the singlet ground states of helium-like systems including positronium negative ion, hydrogen negative ion, helium atom, and lithium positive ion, we also show results for systems with nucleus charge around the ionization threshold.

#E-mail: b99202042@ntu.edu.tw
*E-mail: ykho@pub.iams.sinica.edu.tw

## I. INTRODUCTION

It is important to develop correlation measures in order to increase our understanding of correlation effects. Recently, there has been considerable interest in information measures and their applications in atomic and molecular structures. Entropic measures are used to quantify quantum entanglement and correlations. Quantum entanglement plays a central role in development for research areas such as quantum information, quantum computation, quantum cryptography, and quantum teleportation. In our group, we have studied quantum entanglement measures in natural atomic systems that involve two highly correlated indistinguishable spin-1/2 fermions (electrons). We adopted the linear entropy and von Neumann entropy to quantify entanglement [1 - 5]. Now we move to the Shannon entropy [6], which was first introduced by Claude E. Shannon in 1948 to quantify the average amount of information contained in a source and to characterize the probability distribution of samples. The Shannon entropy is defined by the one-electron charge density. Besides being a stronger version of the Heisenberg uncertainty principle of quantum mechanics [7], the Shannon entropy provides a measure of information about the probability distribution and shows the correlation between particles [8 - 10]. Investigations on Shannon information entropy of atomic systems have also been reported in the literature [11 - 24]. It should be mentioned that we [1 - 5] and others [25 - 29] have calculated von Neumann entropy or linear entropy for two-electron atoms to probe correlation and entanglement. In general, Shannon entropy is a measure of uncertainty of random variables, and it tells us about the localization or delocalization of these two-electron systems. The von Neumann entropy, and its linear approximation, the linear entropy, are measures of correlation in these systems.

In our present work, we calculate the Shannon information entropy in position space for two-electron atomic systems, such as positronium negative ion, hydrogen negative ion, helium atom and lithium positive ion, so as to investigate correlations of charge density distribution between the two electrons. In order to take correlation effects into account, we represent the atomic states with Hylleraas-type wave functions. Once the wave function is obtained, we conduct numerical integrals on the one-electron density distribution to calculate the Shannon entropy. Besides reporting the results for ground states, we also report results for model ions when the charge of the nucleus is decreased from $Z = 1$, the case for hydrogen ion, to the critical charge $Z_{cr}$, at which value the two-electron ion starts to becoming unbound.

## II. THEORETICAL METHOD

The non-relativistic Hamiltonian (in atomic units) describing the helium-like atom/ions is

$$H = -\frac{1}{2}\nabla_1^2 - \frac{1}{2}\nabla_2^2 - \frac{Z}{r_1} - \frac{Z}{r_2} + \frac{1}{r_{12}}, \qquad (1)$$

where 1 and 2 denote the two electrons 1, 2, and $r_{12}$ is the relative distance between the particle 1 and 2. The Hamiltonians with $Z = 1, 2, 3$ are for the hydrogen negative ion, the helium atom and the lithium positive ion respectively. Atomic units in energy are used in the present work.

For *S*-states we use Hylleraas-type wave functions to describe the coordinate part of the system,

$$\Psi_{kmn} = \sum_{kmn} C_{kmn} \left( \exp[-\alpha r_1 - \beta r_2] r_{12}^k r_1^m r_2^n + (1 \leftrightarrow 2) \right), \qquad (2)$$

with $k+m+n \leq \omega$, and $\omega, k, m,$ and $n$ are positive integers or zero. As for positronium negative ions, Ps⁻, similar Hamiltonian and wave functions can be found in an earlier publication [5].

Shannon information entropy for the two-electron systems with electronic charge distribution is given by

$$S = -\int_0^\infty \rho(r) \ln \rho(r) 4\pi r^2 dr, \qquad (3)$$

where the function $\rho(r)$ is the one-electron charge density defined as

$$\rho(r_1) = \frac{2\pi}{r_1} \int_0^\infty \int_{|r_1-r_2|}^{r_1+r_2} |\Psi|^2 \, r_2 dr_2 r_{12} dr_{12}. \qquad (4)$$

The integral in Eq. (3) is solved by the numerical Gauss-Laguerre quadrature integration. The convergence of the numerical calculation is tested by the normalization condition:

$$\int_0^\infty \rho(r) 4\pi r^2 dr = 1 \; . \qquad (5)$$

Eq. (3) shows the expression for the Shannon entropy, and it usually is for bound states. Here, when $Z$ is decreased from $Z > Z_{cr}$ to $Z < Z_{cr}$, the ground state of the two-electron ion becomes a shape resonance, a quasi-bound state (more discussion is given later in the text). While such a quasi-bound state is, straightly speaking, unbound, but its inner part of probability density function has a huge amplitude, in a manner similar to that of a bound state, and as such, it is the underline reason that reasonably accurate results could be obtained for resonance calculations (energy and sometimes width) by using the so-called $L^2$-type (bound-state type) wave functions. Here, as was in Ref. [10], we use Eqs. (3), (4) and (5) (bound-state type integrals) and employing bound-state type wave functions for calculations of properties for quasi-bound resonance states, and we believe that our results are reasonably accurate. As for how quasi-bound resonance states can be calculated by using $L^2$-type wave functions in the stabilization method or in complex-scaling method, readers are referred to references [30, 31].

## III.   CALCULATIONS AND RESULTS

### (A) Shannon entropy for H⁻, He, Li⁺ and Ps⁻

In this work, we calculate Shannon entropy in position space (SEPS) for the ground states of helium-like systems including Ps⁻, H⁻, He, and Li⁺. The ground state wave functions of the isoelectronic systems are constructed by the highly-correlated Hylleraas type function. As the value of $\omega$ in eq. (2) increases, the size of the basis set gets larger and the wave functions would be able to provide better description for the ground states. For the ground state of a given system, by systematically changing the value of $\omega$ we have examined the optimized energy for that state, and with which the Shannon entropy in position space for such state was calculated. In Table 1, we show the convergence of the optimized ground state energy in terms of $\omega$, and hence of $N$. With the ascending number of bases for wave functions, the energies of the isoelectronic analogs converge to at least eight digits after the decimal. The energies are also quite consistent with other accurate results in Ref. [32, 33]. In Table 2, we show the convergence of Shannon entropy in terms of the size of basis set. The Shannon entropy of He and Li⁺ converge better as a consequence of the more compact distribution of the wave functions. We also present a comparison of our results with those in Ref. [19], and it shows that agreement is better for larger $Z$ values than for smaller $Z$. As for Ps⁻, our result for Shannon entropy in position space is the first reported in the literature, to the best of our knowledge. In comparing with hydrogen negative ion, the larger SEPS value in Ps$^-$ shows that the H$^-$ ion is more localized than the Ps$^-$ ion.

In order to shed light on the interpretation of our present results, we also calculate the expectation values of $r_1$ and of $r_{12}$ for the abovementioned two-electron atoms/ions, and the results are shown in Table 3. From these results, it is shown that when $Z$ is increased from $Z=1$ (H$^-$) to $Z=3$ (Li$^+$), the size of the two-electron system shows a decreasing trend with the latter ion becomes more compact. Our results also show that as $Z$ increases, the value of SEPS decreases, indicating the system is more localized for higher $Z$. This is consistent with our decreasing value for the Shannon entropy in position space (more localized) for increasing $Z$. As for H$^-$ and Ps$^-$, both systems have $Z=1$, but the latter ion has a mass polarization term (see, for example, Ref. [34]) that would push the energy up by almost a factor of two. As for the Ps$^-$ ion as compared with that for the H$^-$ ion, the results shown in Table 3 indicate that Ps$^-$ is a more loosely bound ion, implying the system is more delocalized and has a larger value of Shannon entropy in position space.

### (B) Shannon entropy around the critical charge region

Next, we continue to examine the trend of the Shannon entropy in position space around the ionization threshold when $Z$ is decreased from $Z = 1.0$ to $Z = Z_{cr}$, at which value the two-electron ion starts to becoming unbound. The numerical values of the critical charge have been determined for this two-electron ion [35, 36, 37], with the latest entry in the literature [37] providing at least 16 digits after the decimal for accuracy. It was concluded that when the charge of the positively-charged particle is reduced below $Z_{cr}$, the bound state of such a model ion would become a shape resonance and the energy becomes complex [38, 39, 40, 41]. While the real part of such complex energy is related to the resonance position lying above the ionization threshold, the imaginary part of this complex pole is related to the resonance width. The width is inverse proportional to the lifetime of this quasi-bound resonance state, a result of the uncertainty principle. As we will see later in the text that such energy spread for the resonance width is somehow related to the 'delocalization' character for these shape resonances. An investigation of Shannon entropy around the region of critical charge was carried out in Ref. [10]. In Fig. 1 we show our results for SEPS from $Z = 1.0$ to about $Z = 0.88$. When $Z$ is larger than the critical value of about 0.911028, the system remains bound, and the wave functions are obtained based on the usual Rayleigh-Ritz variational bound principle and the energy-optimized wave functions are then used to calculate Shannon entropy as described above. Therefore, in this region with $Z > Z_{cr}$, the convergence for Shannon entropy in position space is very much similar to those shown in Tables 1 and 2. Table 4 shows some selected SEPS values, obtained with $N=444$ terms in the wave functions of Eq. (2), for $Z = 0.92$ to 1.0 in an interval of 0.01, and with these $Z$ values the systems remain bound. For completeness, Table 4 also shows some selected SEPS results for $Z < Z_{cr}$.

For the region when $Z < Z_{cr}$, the situation becomes more complicate, as the bound state now turns into a shape resonance [38 - 41]. As a shape resonance lies in the scattering continuum, the usual Rayleigh-Ritz bound principle to its energy is no longer valid, so care must be taken to choose the wave functions for calculations of Shannon entropy. Here, we take the character of shape resonances into account, and employ a strategy to set the non-linear parameters in wave functions accordingly. In Ref [40] the complex resonance energies were determined using the complex scaling method [31, 42, 43, 44]. Again, our results shown in Fig. 1 for the region when $Z < Z_{cr}$ are obtained using $N = 444$ terms. It is seen that once the nuclear charge is decreased below the critical value, the SEPS starts to increase in a rapid but moderate manner, and eventually it reaches to a saturated value and stay almost constant when the system is away from the shape resonance region. We have further observed that our moderate and gradual increase in SEPS is not a step-like increase as reported in Ref. [10]. This implies that while for a bound state when $Z > Z_{cr}$ the system exhibits the most localized behavior, but when the bound state turns into a shape resonance, the two-electrons will stay 'quasi-bound' for the duration of its lifetime (inverse proportional to the resonance width), before an electron tunnels out of the potential barrier. As a result, the system does not immediately turn delocalized completely, but still exhibits some localized behaviors in the shape resonance region. This explains our findings that when $Z < Z_{cr}$ the Shannon entropy would increase moderately instead of a step-like behavior. It should also be mentioned that in our present work, we have not extended calculations beyond that for $Z = 0.88$, as at such $Z$ values the wave functions would be very much defused, and convergence would be more difficult to achieve. It should also be mentioned that a study of the wave function at the critical charge point was reported in [45], and a plot for some conditional probability density functions around the critical charge region was given in Ref. [10]. But an extensive investigation for details of wave functions around the critical charge should be warranted for a future study and is outside the scope of our present work.

## IV. Summary and Conclusion

By employing highly correlated Hylleraas functions, we have carried out calculations of Shannon information entropy in position space, $S_r$, for two-electron atomic systems such as Ps$^-$, H$^-$, He and Li$^+$. By systematic examination of convergence behaviors when different expansion lengths in the wave functions are used, we are able to provide benchmark values for Shannon entropy in position space in the above-mentioned two-electron atomic systems, with the result for Ps$^-$ being the first reported in the literature. In the future, it would be worthwhile to calculate Shannon information entropy in momentum space, $S_p$, for theses two-electron systems using correlated Hylleraas functions. But the challenging aspect for this approach is that we need to carry out momentum transfer to transform the Hylleraas wave functions from position space to momentum space [46, 47]. The wave functions can then be used for calculations of $S_p$, and the results can be combined with $S_r$ to test the entropic uncertainty principle [7] for these two-electron atomic systems in three dimensional space, with $S_T$ being the entropy sum, as

$$S_T = S_r + S_p \geq 3 (1 + \ln \pi) \quad . \tag{6}$$

The entropy sum can also be used to test a stronger version of Heisenberg's uncertainty principle for an $N$-electron system [7, 48], with

$$S_T \geq 3 N (1 + \ln \pi) - 2N \ln N \quad . \tag{7}$$

Eq. (6) involves calculations of Shannon entropy in momentum space, $S_p$, a quantity we are at present not in a position to calculate. $S_p$ is a measure of momentum distribution for the system. In momentum space, when interaction becomes weaker as $Z$ decreases, the motions of the particles are lesser chaotic, which means that the uncertainty in momentum becomes smaller, leading to the decrease of Shannon entropy in momentum space. Qualitatively, we might expect that Shannon entropy in momentum space would decrease in a rapid fashion at the critical point, showing the decrease of movement of the electrons as $Z$ decreases further, in an opposite fashion as that of the trend in the position space. Nevertheless, it still needs actual calculations to substantiate such a conjecture. It would be of interest to calculate Shannon entropy in momentum space around the critical point of $Z$ for such two-electron ions.

In our present work we also present an investigation on the ground states of these two-electron ions by changing the $Z$ value from $Z = 1.0$ to the region near $Z = Z_{cr}$ at which value this two-electron system starts becoming unbound. For $Z > Z_{cr}$, as the upper-bound principle to their ground states energies applies, we have established benchmark values for Shannon information entropy in position space in this $Z$-region. For $Z < Z_{cr}$, by taking into account of the phenomenon that the bound state would turn into a shape resonance, we have found that the Shannon entropy in position space would increase in a moderate manner, showing a mixture of localization and delocalization behaviors for such shape resonances. It is hope that our findings would stimulate further investigations on such an intrigue behavior for these highly correlated two-electron systems. As for von Neumann entropy, there seems to be no published work, to the best of knowledge, using highly correlated Hylleraas functions to investigate the unbound states (or quasi-bound shape resonances) for the two-electron ions when $Z < Z_{cr}$. In passing, we are aware of an attempt to investigate such phenomenon using the $S$-wave model (keeping just the leading monopole term in the expansion of the electron-electron interacting potential operator) for such two-electron systems [49]. Quantification of von Neumann entropy and linear entropy for doubly excited Feshbach resonances in two-electron atoms has recently been carried out [50, 51]. It would be of interest in the future to quantify von Neumann entropy for shape resonance states in the region when $Z < Z_{cr}$ using highly correlated wave functions and keeping the full Hamiltonian for these two-electron systems.

## ACKNOWLEDGEMENT

Our work has been supported by the Ministry of Science and Technology of Taiwan.

Table 1. The ground state energy for the systems in terms of the number of basis set $N$ in Hylleraas functions.

| $\omega$ | $N$ | Ps$^-$ | H$^-$ | He | Li$^+$ |
|---|---|---|---|---|---|
| 9 | 125 | -0.2620049976 | -0.5277508541 | -2.9037243542 | -7.2799133884 |
| 10 | 161 | -0.2620050506 | -0.5277509652 | -2.9037243682 | -7.2799134037 |
| 11 | 203 | -0.2620050607 | -0.5277509892 | -2.9037243734 | -7.2799134090 |
| 12 | 252 | -0.2620050665 | -0.5277509986 | -2.9037243754 | -7.2799134111 |
| 13 | 308 | -0.2620050683 | -0.5277510107 | -2.9037243762 | -7.2799134119 |
| 14 | 372 | -0.2620050693 | -0.5277510137 | -2.9037243767 | -7.2799134123 |
| 15 | 444 | -0.2620050698 | -0.5277510152 | -2.9037243768 | -7.2799134125 |
| Ref. [32] | | -0.2620050702 | -0.5277510165 | -2.9037243770 | |
| Ref. [33] | | | -0.5277510165 | -2.9037243770 | -7.2799134127 |

Table 2. The Shannon entropy in position space for the systems in terms of the number of basis set N in Hylleraas functions.

| $\omega$ | $N$ | Ps⁻ | H⁻ | He | Li⁺ |
|---|---|---|---|---|---|
| 9 | 125 | 7.9552380 | 5.837120 | 2.7051024 | 1.2552724 |
| 10 | 161 | 7.9552832 | 5.837153 | 2.7051027 | 1.2552725 |
| 11 | 203 | 7.9552923 | 5.837162 | 2.7051028 | 1.2552725 |
| 12 | 252 | 7.9552978 | 5.837174 | 2.7051028 | 1.2552725 |
| 13 | 308 | 7.9552995 | 5.837172 | 2.7051028 | 1.2552726 |
| 14 | 372 | 7.9553005 | 5.837172 | 2.7051028 | 1.2552726 |
| 15 | 444 | 7.9553009 | 5.837173 | 2.7051028 | 1.2552726 |
| Ref. [19] | | | 5.6115 | 2.7117 | 1.2944 |

Table 3. Expectation values of $r_1$ and $r_{12}$ for the ground states of two-electron systems.

| | Ps⁻ | H⁻ | He | Li⁺ |
|---|---|---|---|---|
| $<r_1>$ Present | 5.4896316973 | 2.7101771334 | 0.9294722930 | 0.5727741496 |
| Others | 5.4896332523[a] | 2.7101782784[b] | 0.9294722948[b] | 0.5727741499[b] |
| $<r_{12}>$ Present | 8.5485776017 | 4.4126922528 | 1.4220702521 | 0.8623153749 |
| Others | 8.5485806551[a] | 4.4126944979[b] | 1.4220702555[b] | 0.8623153754[b] |

[a][32]; [b][52]

Table 4. The Shannon entropy in position space for two-electron systems with nucleus charge Z, below and above the critical value $Z_{cr}$ = 0.911, ranging from 0.88 to 1.00.

| Z | Energy | $S_r$ |
|---|---|---|
| 0.880 | -0.3837953968 | 10.115539 |
| 0.890 | -0.3928805004 | 10.055493 |
| 0.895 | -0.3975127443 | 9.804300 |
| 0.900 | -0.4025301585 | 7.899564 |
| 0.910 | -0.4137989542 | 6.849090 |
| 0.920 | -0.4254852567 | 6.620761 |
| 0.930 | -0.4374512977 | 6.468422 |
| 0.940 | -0.4496690375 | 6.346169 |
| 0.950 | -0.4621246954 | 6.241010 |
| 0.960 | -0.4748098319 | 6.147163 |
| 0.970 | -0.4877187106 | 6.061483 |
| 0.980 | -0.5008471781 | 5.982044 |
| 0.990 | -0.5141920954 | 5.907571 |
| 1.000 | -0.5277510151 | 5.837173 |

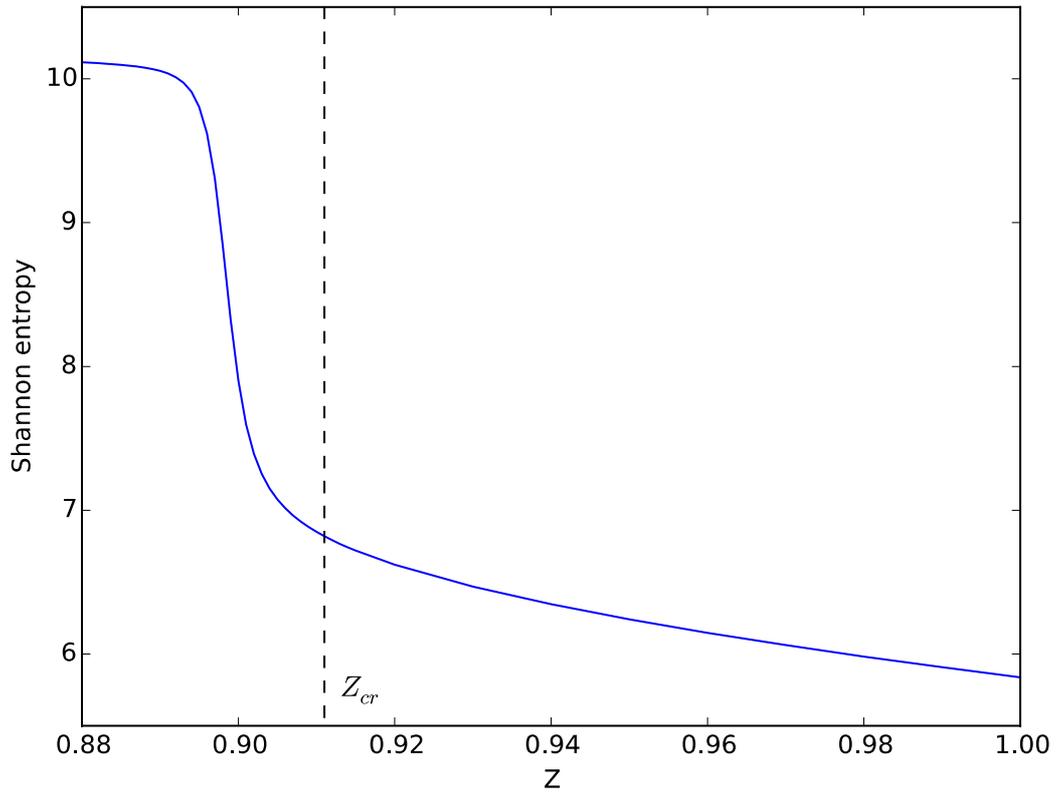

Figure 1. Shannon entropy in position space for systems around the critical nuclear charge $Z_{cr}$ below which the system becomes unbound, and turns into a shape resonance.